# Characterising Linear Spatio-Temporal Dynamical Systems in the Frequency Domain

Hua-Liang Wei, Stephen A. Billings

Department of Automatic Control and Systems Engineering
The University of Sheffield
Mappin Street, Sheffield,
S1 3JD, UK

w.hualiang@sheffield.ac.uk, s.billings@sheffield.ac.uk

**Abstract:** A new concept, called the spatio-temporal transfer function (STTF), is introduced to characterise a class of linear time-invariant (LTI) spatio-temporal dynamical systems. The spatio-temporal transfer function is a natural extension of the ordinary transfer function for classical linear time-invariant control systems. As in the case of the classical transfer function, the spatio-temporal transfer function can be used to characterise, in the frequency domain, the inherent dynamics of linear time-invariant spatio-temporal systems. The introduction of the spatio-temporal transfer function should also facilitate the analysis of the dynamical stability of discrete-time spatio-temporal systems.

## 1. Introduction

It is known that linear time-invariant (LTI) dynamical systems can be described by ordinary differential or difference equations, which can easily be transformed to a compact form of either continuous-time or discrete-time transfer functions, by means of the Fourier, Laplace or Z transforms, under some assumptions on the system initial conditions. Transfer functions, which can be used to uniquely characterize LTI systems, are a useful tool for the analysis, design and control of such systems.

Spatio-temporal dynamical systems are complex systems where the system states evolve spatially as well as temporally. Unlike classical temporal systems where the current output is a function of previous inputs and outputs only in time, the output of a spatio-temporal system depends not only on past values in time but also values at different spatial locations (Coca and Billings, 2001; Billings and Coca, 2002). Spatio-temporal phenomena exist widely in biology, chemistry, ecology, geography, medicine, physics, and sociology (Kaneko, 1986; Jahne, 1993; Silva et al., 1997; Bascompte and Sole, 1998; Czaran, 1998; Billings et al., 2006; Guo et al., 2006). A commonly used theoretical description of spatio-temporal systems is often given in terms of partial differential or difference equations (PDE's) (Ames, 1992; Strikwerda, 1989); Other representations, for example coupled map lattice (CML) models (Kaneko, 1986; Coca and Billings, 2001; Billings and Coca, 2002; Billings et al., 2002) are also be applied to approximate such systems. It has been noticed that a wide class of spatio-temporal dynamical systems and phenomena in the real world can be characterised or approximately described by linear PDE's (Roesser, 1975; Polianin, 1992). In the literature, linear partial differential equations, as a class of continuous-time distributed parameter models, have been extensively studied and several approaches have been developed to analytically or numerically solve these kind of equations.

Integral transformations, including the Fourier and Laplace transforms, play an important role for the analysis of multivariable systems, which are either linear or nonlinear (Lubbock and Bansal, 1969; Parente,1970; Chua and Ng, 1979; Rugh, 1981; Churchill, 1987; Zhang and Billings,1993,1994; Augusta and Hurák, 2006). These transformations are one of a set of the basic tools which can be used to analytically solve linear partial differential equations (Duffy, 2004; Evans et al., 2000; Rabenstein and Trautmann, 2002).

Inspired by the concept of the transfer function for classical LTI dynamical systems, this study aims to introduce a parallel concept, depicted by a multivariable function, that can be used to characterize, in the frequency domain, the underlying dynamics of dynamical LTI spatio-temporal systems described by linear PDE's. The newly introduced function will be called the *spatio-temporal transfer function* (STTF), which is derived by using multivariable Laplace or Fourier transforms. The concept of the STTF is, as far as we are aware, a totally new way of studying spatio-temporal systems in the frequency domain. As will be seen, STTF plays an important role for characterizing LTI spatio-temporal systems. In addition, the introduction of STTF can facilitate the analysis of the dynamical stability of linear partial difference equations for some continuous-time spatio-temporal systems.

## 2. The spatio-temporal transfer function

For simplicity, the case of 2-D linear partial differential equations is considered as an example to illustrate the concept of the STTF, but note that similar derivations can directly be extended to cases of arbitrary *n*-dimensional problems where *n*>2. A wide class of spatio-temporal systems can be described by linear partial differential equations of the form below

$$a\phi_{xx} + b\phi_{xy} + c\phi_{yy} + d\phi_x + e\phi_y + g\phi(x,y) = u(x,y) \tag{1}$$

where $a, b, c, d, e$ and $g$ are constants, $x$ and $y$ are two independent variables, $\phi(x,y)$ is a real-valued function that is usually known, and $u(x,y)$ is a known input (control) function; the functions $\phi_x$, $\phi_y$, $\phi_{xx}$, $\phi_{xy}$ and $\phi_{yy}$ are partial derivatives defined by following normal convention, for example, $\phi_x = \partial\phi/\partial x$ and $\phi_{xy} = \partial^2\phi/\partial x\partial y$.

Let $\Phi(s_1, s_2)$ be the Laplace transform of the functions $\phi(x, y)$, and $R_\phi$ be the relative region of convergence (ROC). The Laplace transform pair associated with the function $\phi(x, y)$, over $R_\phi$, is

$$\Phi(s_1, s_2) = \int_{-\infty}^{\infty} \int_{-\infty}^{\infty} \phi(x, y) e^{-(xs_1 + ys_2)} dx dy \tag{2}$$

$$\phi(x, y) = \frac{1}{(2\pi j)^2} \int_{\sigma_1 - \infty}^{\sigma_1 + \infty} \int_{\sigma_2 - \infty}^{\sigma_2 + \infty} \Phi(s_1, s_2) e^{xs_1 + ys_2} ds_1 ds_2 \tag{3}$$

where $j = \sqrt{-1}$, the real-valued numbers $\sigma_1$ and $\sigma_2$ can be chosen within the region of convergence (ROC) of the function $\Phi(s_1, s_2)$. Under the assumption that $\lim_{x \to \pm\infty} \phi(x, y) = 0$, the partial derivative function $\phi_x$ can be obtained by differentiating both sides of the synthesis equation (3) as below

$$\phi_x = \frac{\partial \phi}{\partial x} = \frac{1}{(2\pi j)^2} \int_{\sigma_1 - \infty}^{\sigma_1 + \infty} \int_{\sigma_2 - \infty}^{\sigma_2 + \infty} s_1 \Phi(s_1, s_2) e^{xs_1 + ys_2} ds_1 ds_2 \tag{4}$$

Hence, from the definition (2), the Laplace transform of the function $\phi_x$ is

$$\Phi_{(x)}(s_1, s_2) = L[\phi_x(x, y)] = s_1 \Phi(s_1, s_2) \tag{5}$$

Similarly, with the assumption that $\lim_{y \to \pm\infty} \phi(x, y) = 0$, $\lim_{x \to \pm\infty} \phi_x(x, y) = 0$, $\lim_{y \to \pm\infty} \phi_y(x, y) = 0$, $\lim_{x \to \pm\infty} \phi_y(x, y) = \lim_{y \to \pm\infty} \phi_x(x, y) = 0$, the Laplace transform of the partial derivative functions $\phi_y$, $\phi_{xx}$, $\phi_{xy}$ and $\phi_{yy}$ are: $\Phi_{(y)}(s_1, s_2) = s_2 \Phi(s_1, s_2)$, $\Phi_{(xx)}(s_1, s_2) = s_1^2 \Phi(s_1, s_2)$, $\Phi_{(xy)}(s_1, s_2) = s_1 s_2 \Phi(s_1, s_2)$, and $\Phi_{(yy)}(s_1, s_2) = s_2^2 \Phi(s_1, s_2)$, respectively.

Let $U(s_1, s_2)$ be the Laplace transforms of the functions $u(x, y)$, and $R_u$ be the relative ROC. Taking the Laplace transform for both sides of (1), yields,

$$P(s_1, s_2) \Phi(s_1, s_2) = U(s_1, s_2) \tag{6}$$

where the relative ROC contains $R_\phi \cap R_u$, and

$$P(s_1, s_2) = as_1^2 + bs_1 s_2 + cs_2^2 + ds_1 + es_2 + g \tag{7}$$

Thus, the spatio-temporal transfer function (STTF), relative to the PDE (1), can be defined as

$$G(s_1, s_2) = \frac{\Phi(s_1, s_2)}{U(s_1, s_2)} = \frac{1}{P(s_1, s_2)} \tag{8}$$

Consequently, for arbitrary input function $u(x, y)$, the output response function of (1) is

$$\Phi(s_1, s_2) = G(s_1, s_2) U(s_1, s_2) = \frac{U(s_1, s_2)}{P(s_1, s_2)} \tag{9}$$

The polynomial $P(s_1, s_2)$ given by (7) is the associated characteristic polynomial of the system described by (1), and $P(s_1, s_2) = 0$ is the relevant characteristic equation. Following traditional routines for the determination of characteristic polynomials for LTI ODE's, a simple way to derive the characteristic polynomial $P(s_1, s_2)$ here, is to set $u(x,y) = e^{s_1 x + s_2 y}$ and $\phi(x,y) = \Phi(s_1, s_2)e^{s_1 x + s_2 y}$, and then substitute these expressions into the partial differential equation (1), the Laplace transform $\Phi(s_1, s_2)$ and the characteristic polynomial $P(s_1, s_2)$ can then be solved by performing some algebra.

For a given system, the spatio-temporal transfer function (8) is unique, and the response function (9) is thus also deterministic once the input has been given. By setting $s_1 = j\omega_1$ and $s_2 = j\omega_2$, the Laplace transform pair (2) and (3) will become to the Fourier transform pairs of the relevant functions, and the resultant output frequency response function can then be used to analyze the inherent frequency property of given spatio-temporal systems.

### 3. Finite difference schemes and the discrete-time spatio-temporal transfer function

Finite difference (FD) schemes are a simple approach to solve differential equations by means of differencing methods (Ames, 1992; Strikwerda, 1989). Taking the partial differential equation (3) as an example, finite difference schemes initially define a grid of points in the $(x, y)$ plane. Let $h$ and $k$ be positive numbers, the grid will then be the points $(x_n, y_m) = (nh, mk)$ for arbitrary integers $n$ and $m$. Denote the value of the function $\phi(x,y)$ at the grid point $(x_n, y_m)$ by $\phi[n,m]$, then the partial derivatives can be approximated using the central difference method (Ames, 1992) below

$$\phi_x[n,m] = \phi_x(nh, mk) \approx \frac{\phi[n+1,m] - \phi[n-1,m]}{2h} \tag{10}$$

$$\phi_y[n,m] = \phi_y(nh, mk) \approx \frac{\phi[n,m+1] - \phi[n,m-1]}{2k} \tag{11}$$

$$\phi_{xx}[n,m] = \phi_{xx}(nh, mk) \approx \frac{\phi[n+1,m] - 2\phi[n,m] + \phi[n-1,m]}{h^2} \tag{12}$$

$$\phi_{yy}[n,m] = \phi_{yy}(nh, mk) \approx \frac{\phi[n,m+1] - 2\phi[n,m] + \phi[n,m-1]}{k^2} \tag{13}$$

$$\phi_{xy}[n,m] = \phi_{xy}(nh, mk) \approx \frac{\phi[n+1,m-1] - \phi[n-1,m-1] - \phi[n+1,m+1] + \phi[n-1,m+1]}{4hk} \tag{14}$$

By applying the 2-D Z-transform (Dudgeon and Mersereau, 1984) to equations from (10) to (14), and then substituting the relevant results to (1), yields

$$P(z_1, z_2)\Phi(z_1, z_2) = U(z_1, z_2) \tag{15}$$

where

$$P(z_1, z_2) = c_1(z_1 z_2^{-1} - z_1^{-1} z_2^{-1} - z_1 z_2 + z_1^{-1} z_2) + c_2 z_1 + c_3 z_2 + c_4 z_1^{-1} + c_5 z_2^{-1} + c_6 \tag{16}$$

is the associated characteristic polynomial, and the relative coefficients are given below

$$c_1 = \frac{b}{4hk}, \quad c_2 = \frac{a}{h^2} + \frac{d}{2h}, \quad c_3 = \frac{c}{k^2} + \frac{e}{2k}, \quad c_4 = \frac{a}{h^2} - \frac{d}{2h}, \quad c_5 = \frac{c}{k^2} - \frac{e}{2k}, \quad c_6 = g - \frac{2a}{h^2} - \frac{2c}{k^2},$$

From (15), the discrete-time spatio-temporal transfer function of the system (1) is given by

$$H(z_1, z_2) = \frac{\Phi(z_1, z_2)}{U(z_1, z_2)} = \frac{1}{P(z_1, z_2)} \tag{17}$$

Similar to the spatio-temporal transfer function (8), the transfer function (17) provides a representation for given LTI parameter distributed systems. Note that for a given system, the discrete-time STTF (17) may not be unique, because the derivation of this function is relative to the finite difference scheme employed. Different difference schemes will lead to different transfer functions. However, once the difference schemes have been determined, the resultant transfer function will be unique.

One advantage of the introduction of the discrete-time STTF is that this function can facilitate the analysis of the dynamical stability of the spatio-temporal systems. With regard to the discrete-time transfer function (17), the concept of two stabilities is usually distinguished: the numerical stability that is relative to different difference schemes, and the dynamical stability that is determined by the underlying dynamics of the systems. Generally, numerical stability is independent of the inherent dynamics of the systems. Detailed discussions on numerical stability analysis, relative to different difference schemes, can be found in Ames (1992) and Strikwerda (1989). This study touches upon dynamical stability analysis and BIBO (bounded input and bounded output) stability will be considered.

Whilst it is difficult to analyze the stability of given spatio-temporal systems directly using the STTF (8) because of the lack of existing tools, the dynamical stability analysis for the STTF (17) is more tractable and several stability theorems are available, see for example Huang (1972), Shanks et al. (1972), Justice and Shanks (1973), Anderson and Jury (1974), Strintzis (1977), and Dudgeon and Mersereau (1984) (most of the earliest work on stability analysis can be found in this book). Some recent results on stability analysis have been reported in Bistritz (1999, 2004), Curtin and Saba (1999), Damera-Venkata et al. (2000), and Mastorakis (2000).

## 4. Numerical examples

This section provides two examples to illustrate the application of the spatio-temporal transfer function for LTI spatio-temporal systems.

*4.1 The wave equation*

The wave equation is given below

$$\alpha^2 \phi_{xx} - \phi_{tt} = 0, \text{ for } -\infty \leq x \leq \infty \text{ and } t > 0 \tag{18}$$

where $\alpha$ is a positive number. In (7), by setting $a = \alpha^2$, $c = -1$, and $b = d = e = g = 0$, the characteristic equation of the system (18) can be calculated to be

$$P(s_1, s_2) = \alpha^2 s_1^2 - s_2^2 = 0 \tag{19}$$

Let $s_1 = j\omega_1$, $s_2 = j\omega_2$, then from (19)

$$\omega_2 = \pm \alpha \omega_1 \tag{20}$$

Equation (20) clearly shows that the temporal frequency (in the time direction) in the wave system (18) is $\alpha$ times the associated spatial frequency (in the spatial direction). The relationship (20), revealed by the STTF approach, is coincident with the result obtained via analytical approaches. For example, given the initial condition: $\phi(x,0) = \varphi(x)$, $\phi_t(x,0) = \psi(x)$ for $-\infty \leq x \leq \infty$, the well-known d'Alembert's formula, relative to the Cauchy problem, states that

$$\phi(x,t) = \frac{1}{2}[\varphi(x-\alpha t) + \varphi(x+\alpha t)] + \frac{1}{2\alpha}\int_{x-\alpha\tau}^{x+\alpha\tau}\psi(\tau)d\tau \tag{21}$$

This formula clearly indicates that the relationship between the temporal frequency and the spatial frequency is given by (21), and this is independent of the choice of the initial condition functions $\varphi(x)$ and $\psi(x)$.

Now, consider the dynamical stability of the discrete-time spatio-temporal transfer function of the wave equation (18). From (16), the characteristic polynomial for the wave equation is given by

$$P(z_1, z_2) = \frac{\alpha^2}{h^2}(z_1 + z_1^{-1} - 2) - \frac{1}{k^2}(z_2 + z_2^{-1} - 2) \tag{22}$$

The stability condition of $P(z_1, z_2)$ is equivalent to that of the polynomials below

$$Q_0(z_1, z_2) = (z_2 + z_2^{-1} - 2) - \lambda(z_1 + z_1^{-1} - 2) \tag{23}$$

and

$$Q(z_1, z_2) = z_1(1 - z_2)^2 - \lambda z_2(1 - z_1)^2 \tag{24}$$

where $\lambda = \alpha^2 k^2 / h^2$. From relevant theorems for stability analysis (Huang, 1972; Anderson and Jury, 1974), to demonstrate stability of the two-variable polynomial $Q(z_1, z_2)$, we need to check that

$$Q(z_1, 0) \neq 0 \text{ for } |z_1| \leq 1, \tag{25}$$

and

$$Q(z_1, z_2) \neq 0 \text{ for } |z_1| = 1, |z_2| \leq 1, \tag{26}$$

It is clear that (25) holds if and only if $z_1 = 0$. Now consider the condition (26). Let $z_1 = e^{j\theta_1}$, $z_2 = \beta e^{j\theta_2}$, with $0 < \beta \leq 1$, $0 \leq \theta_1 \leq 2\pi$, and $0 \leq \theta_2 \leq 2\pi$. Clearly, $|z_1| = 1, |z_2| \leq 1$, and if $Q(z_1, z_2) = 0$ then $Q_0(z_1, z_2) = z_1^{-1} z_2^{-1} P(z_1, z_2) = 0$. From $Q_0(z_1, z_2) = 0$,

$$[(\beta + \beta^{-1})\cos(\theta_2) - 2 + j(\beta - \beta^{-1})\sin(\theta_2)] - \lambda[2\cos(\theta_1) - 2] = 0 \tag{27}$$

Thus, by respectively equating the real and imaginary terms in (27), it can be obtained that

$$(\beta + \beta^{-1})\cos(\theta_2) - 2 = 2\lambda[\cos(\theta_1) - 1] \tag{28}$$

$$(\beta - \beta^{-1})\sin(\theta_2) = 0 \tag{29}$$

Noting that $0 < \beta \leq 1$ and $0 < \lambda \leq 1$ (required by the numerical stability of the relevant finite difference scheme), it can be concluded by combining (28) and (29) that either $\theta_2 = 0$ or $\beta = 1$. If $\theta_2 = 0$, then from (28) $\theta_1 = 0$ and $\beta = 1$; If $\beta = 1$, then $\cos(\theta_2) = 1 - \lambda + \lambda \cos(\theta_1)$.

In conclusion, the characteristic polynomial $P(z_1, z_2)$, relative to the wave equation (18), is marginally stable; the critical points are: i) $z_1 = 0$, $z_2 = 0$, and ii) those defined by $z_1 = e^{j\theta_1}$, $z_2 = e^{j\theta_2}$, satisfying $\cos(\theta_2) = 1 - \lambda + \lambda \cos(\theta_1)$.

*4.2 Poisson's equation*

Consider Poisson's equation given below

$$\phi_{xx} + \phi_{yy} + \phi_{zz} = u(x, y, z), \text{ for } -\infty \leq x, y, z \leq \infty, \tag{30}$$

where the input signal is of the form

$$u(x, y, z) = K e^{-(a_1 x + a_2 y + a_3 z)} \sin(\omega_1' x) \sin(\omega_2' y) \sin(\omega_3' z) \tag{31}$$

and $K, a_1, a_2, a_3$ and $\omega_1', \omega_2', \omega_3'$ are known parameters. The spatio-temporal transfer function of Poisson's equation (34) is

$$G(j\omega_1, j\omega_2, j\omega_3) = \frac{-1}{\omega_1^2 + \omega_2^2 + \omega_3^2} \tag{32}$$

and the output frequency response function of the system driven by the input signal (31) is

$$\Phi(j\omega_1, j\omega_2, j\omega_3) = G(j\omega_1, j\omega_2, j\omega_3) U(j\omega_1, j\omega_2, j\omega_3)$$
$$= \frac{K \omega_1' \omega_2' \omega_3'}{P_0(\omega_1, \omega_2, \omega_3) P_1(\omega_1) P_2(\omega_2) P_3(\omega_3)} \tag{33}$$

where $P_0(\omega_1, \omega_2, \omega_3) = -(\omega_1^2 + \omega_2^2 + \omega_3^2)$ and $P_k(\omega_k) = (a_k^2 + \omega_k'^2 - \omega_k^2) + j2a_k\omega_k$ for $k=1,2,3$. The magnitude and phase (angle) spectra of the output frequency response function are respectively defined as below:

$$|\text{PHY}| = |\Phi| = |\Phi(j\omega_1, j\omega_2, j\omega_3)| = \frac{K\omega_1'\omega_2'\omega_3'}{|P_0(\omega_1, \omega_2, \omega_3) P_1(\omega_1) P_2(\omega_2) P_3(\omega_3)|} \tag{34a}$$

$$\text{Ang(PHY)} = \text{Angle}(\Phi) = \tan^{-1}\left(\frac{\text{Im}(\Phi(j\omega_1, j\omega_2, j\omega_3))}{\text{Re}(\Phi(j\omega_1, j\omega_2, j\omega_3))}\right) \tag{34b}$$

Note that $|P_0(\omega_1, \omega_2, \omega_3)|$ has a peak at $\omega_1 = \omega_2 = \omega_3 = 0$; $|P_k(\omega_k)|$ has peaks at $\omega_k = \pm \omega_k'$, with $k=1,2,3$. The peaks of the magnitude spectrum of the output frequency response function (33) should thus appear at $\omega_k = 0$ and/or $\omega_k = \pm \omega_k'$. The phase spectrum of $P_k^{-1}(\omega_k)$ moves smoothly from $\pi$ to $-\pi$, passing the origin, when $\omega_k$ varies from negative to positive values.

Now, consider the case where the parameters are chosen to be $K=30$, $a_1=0.1$, $a_2=0.2$, $a_3=0.3$, $\omega_1'=1$, $\omega_2'=2$, and $\omega_3'=3$. To graphically illustrate the property of the output frequency response function in the 3-D space, the following scenario was considered: $\omega_3$ was chosen to be a set of fixed values but $\omega_1$ and $\omega_2$ were permitted to vary freely. Numerical results show that when $\omega_3$ is small, the magnitude spectrum at the origin is large. When $\omega_3$ becomes large, however, the magnitude spectrum will be dominated by $\omega_k = \pm\omega'_k$ and the peak at $\omega_k = 0$ becomes invisible. The graph of the output frequency response function (33), corresponding to $\omega_3 = 0.5$, is shown in Fig. 1, where the magnitude and the phase spectrum, along with the relevant contour plots, are presented. Figure 1 clearly shows that for the fixed value $\omega_3 = 0.5$, the magnitude of the output frequency response function (33) has seven peaks at $\omega_1 = \omega_2 = 0$, $\omega_1 = \pm 1$, and $\omega_2 = \pm 2$. This fact is also reflected from the phase spectrum, which varies smoothly from $\pi$ to $-\pi$, passing the origin, when $\omega_k$ varies from negative values ($\omega_k < -\omega'_k$) to positive values ($\omega_k > \omega'_k$) for $k=1,2$.

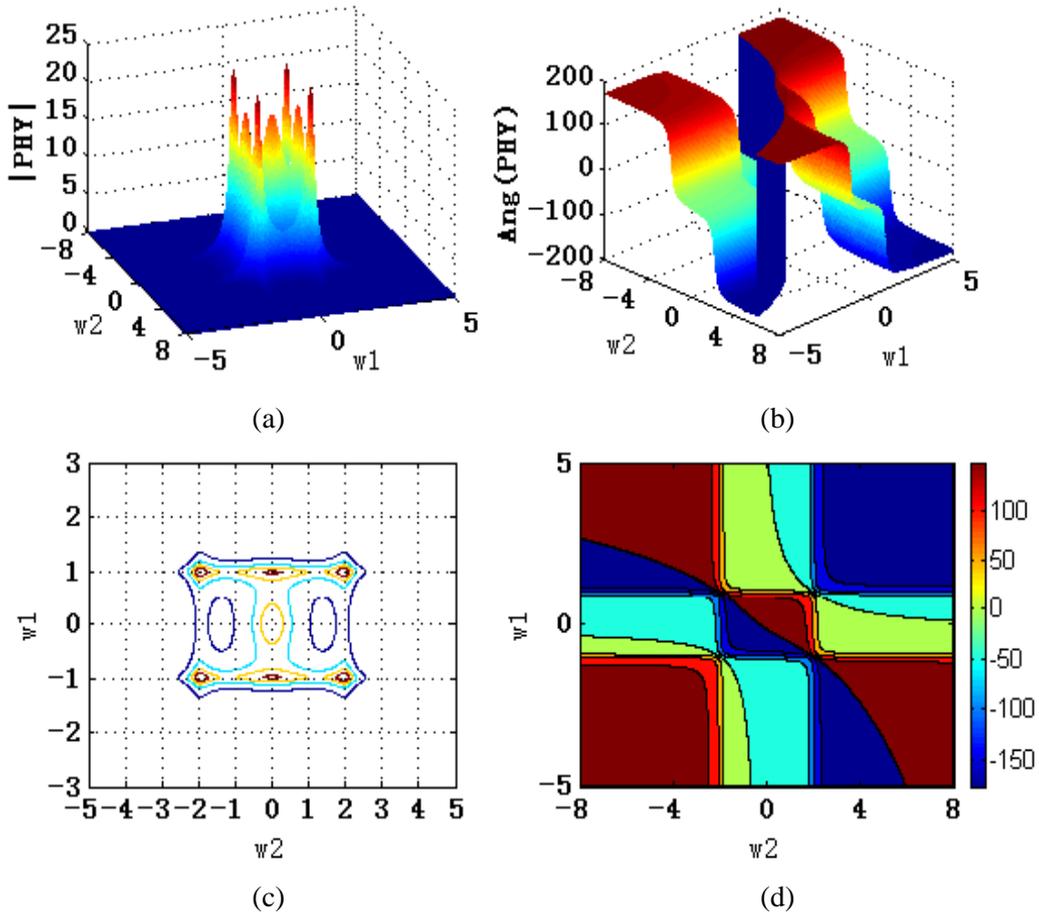

Fig. 1. The magnitude and phase spectra of the output frequency response function $\Phi(j\omega_1, j\omega_2, j\omega_3)$ given by (33), with $\omega_3 = 0.5$. (a) the magnitude spectrum; (b) the phase spectrum; (c) the contour plot of (a); (d) the contour graph of (b).

## 5. Conclusions

Based on the traditional integral transformations, the spatio-temporal transfer function (STTF) has been introduced and applied to analyse, in the frequency domain, the inherent dynamics of a class of spatio-temporal systems. The STTF, along with the relative frequency response function, can be used to reveal the frequency properties of any given LTI spatio-temporal system, because every such system possesses a unique STTF.

By introducing the discrete time STTF, the analysis of the dynamical stability of partial differential equations becomes possible because several existing stability analysis theorems can now be applied to the discrete time STTF. The stability analysis of continuous STTF's has not been investigated in this study. How to analyse the stability of a given LTI spatio-temporal system, directly using the associated continuous multivariable STTF, is a topic for future study.

**Acknowledgements**

The authors gratefully acknowledge that this work was supported by the Engineering and Physical Sciences Research Council (EPSRC), U.K.